\documentclass[reprint,aps,prl,amsmath,amssymb,superscriptaddress, preprintnumbers]{revtex4-2}

\usepackage[utf8]{inputenc}
\usepackage[T1]{fontenc}

\usepackage{amsmath}
\usepackage{mathrsfs}
\usepackage{txfonts}
\usepackage{mathtools}
\usepackage{braket}
\usepackage{tensor}
\usepackage{xcolor}
\usepackage{siunitx}
\usepackage{graphicx}
\usepackage{booktabs}

\usepackage{float}
\usepackage{ulem} 
\usepackage[colorlinks=true,linkcolor=blue,citecolor=blue,urlcolor=blue]{hyperref}

\newcommand{\be}{\begin{equation}}
\newcommand{\ee}{\end{equation}}
\newcommand{\bea}{\begin{align}}
\newcommand{\eea}{\end{align}}

\newcommand{\beq}{\begin{equation}}
\newcommand{\eeq}{\end{equation}}
\newcommand{\beqn}{\begin{eqnarray}}
\newcommand{\eeqn}{\end{eqnarray}}
\newcommand{\bsub}{\begin{subequations}}
\newcommand{\esub}{\end{subequations}}
\newcommand{\bpm}{\begin{pmatrix}}
\newcommand{\epm}{\end{pmatrix}}

\newcommand{\eMax}{\ensuremath{e_{\text{Max}}}}

\newcommand{\ME}[3]{\langle #1 \| #2 \| #3 \rangle}
\newcommand{\Mtwo}{M^{2\nu}}
\newcommand{\Mzero}{M^{0\nu}}
\newcommand{\GT}{\mathrm{GT}}
\newcommand{\safeinclude}[2]{%
  \IfFileExists{#1}{\includegraphics[width=\columnwidth]{#1}}%
  {\fbox{\parbox[c][0.23\textheight][c]{0.95\columnwidth}{\centering #2\\[0.5ex]\texttt{\detokenize{#1}}}}}%
}

\newcommand{\safeincludegraphics}[2][]{%
  \IfFileExists{#2}{\includegraphics[#1]{#2}}%
  {\fbox{\parbox[c][0.23\textheight][c]{0.95\columnwidth}{\centering Missing figure\\[0.5ex]\texttt{\detokenize{#2}}}}}%
}

\begin{document}
\title{\textit{Ab initio} correlations between neutrinoless and two-neutrino double-beta decays in $^{48}$Ca}

\author{X.~Lian} 
\affiliation{College of Physics, Sichuan University, Chengdu 610065, China}

\author{C.~R.~Ding}  
\affiliation{School of Physics and Astronomy, Sun Yat-sen University, Zhuhai 519082, P.R. China}
\affiliation{Guangdong Provincial Key Laboratory of Quantum Metrology and Sensing, Sun Yat-sen University, Zhuhai 519082, P.R. China}

\author{C.~L.~Bai}
\email{bclphy@scu.edu.cn}
\affiliation{College of Physics, Sichuan University, Chengdu 610065, China}

\author{J.~M.~Yao}
\email{yaojm8@sysu.edu.cn}
\affiliation{School of Physics and Astronomy, Sun Yat-sen University, Zhuhai 519082, P.R. China}
\affiliation{Guangdong Provincial Key Laboratory of Quantum Metrology and Sensing, Sun Yat-sen University, Zhuhai 519082, P.R. China}
    \affiliation{Yukawa Institute for Theoretical Physics, Kyoto University, Kyoto, 606-8502, Japan}  

\date{\today}

\begin{abstract}
We develop a novel \textit{ab initio} in-medium no-core configuration-interaction (IM-NCCI) framework for nuclear charge-exchange processes by combining the in-medium similarity renormalization group with chiral nuclear Hamiltonians, and apply it to the $2\nu\beta\beta$ and $0\nu\beta\beta$ decays of $^{48}$Ca. The framework reproduces reasonably the locations of several main resonance peaks in the Gamow--Teller (GT) strength distribution for the $^{48}\mathrm{Ca}\to{}^{48}\mathrm{Sc}$ transition. The cumulative GT strength indicates missing contributions from two-body weak currents, corresponding to an effective quenching factor of $q\simeq0.84$. Incorporating this quenching yields a $2\nu\beta\beta$ nuclear matrix element (NME) in excellent agreement with experiment. Applying the same framework to $0\nu\beta\beta$ decay, and including the contribution from short-range operators, we obtain a total NME of $\Mzero=1.00\text{--}2.02$. Using 34 non-implausible chiral Hamiltonians, we establish from first principles strong linear correlations between the $0\nu\beta\beta$ NME and the NMEs governing $2\nu\beta\beta$ decay and double GT transitions. Combining these correlation relations within the 95\% confidence level with the experimental $2\nu\beta\beta$-decay data yields a constrained prediction of $\Mzero=1.30\text{--}1.65$. This work establishes IM-NCCI as a complementary \textit{ab initio} framework for nuclear weak decays and opens a pathway toward constraining $0\nu\beta\beta$ NMEs in heavier candidate nuclei using experimentally accessible $2\nu\beta\beta$-decay data.

\end{abstract}

\maketitle

\paragraph{Introduction.$-$} Neutrino oscillation experiments have established that neutrinos are massive, providing direct evidence for physics beyond the Standard Model~\cite{Super-Kamiokande:1998,ahmad2001,eguchi2003,Dayabay:2012}. However, the questions about the nature and absolute masses of neutrinos remain unresolved. Neutrinoless double-$\beta$ ($0\nu\beta\beta$) decay provides a unique probe of these questions. It is a second-order weak process in which a nucleus $(A,Z)$ decays into $(A,Z+2)$ with the emission of two electrons and no neutrinos~\cite{Furry:1939,Engel:2017}. Its observation would establish lepton-number violation and have profound implications for the  origin of the matter–antimatter asymmetry in the Universe.

If $0\nu\beta\beta$ decay is dominated by light-Majorana-neutrino exchange, its half-life can be related to the effective Majorana neutrino mass, $\langle m_{\beta\beta}\rangle=\sum_i U^2_{ei}m_i$ through a nuclear matrix element (NME) $\Mzero$, where $m_i$ are the light-neutrino masses and $U_{ei}$ are elements of the  Pontecorvo-Maki-Nakagawa-Sakata
flavor-mixing matrix.  Extracting $\langle m_{\beta\beta}\rangle$ from a measured half-life, or translating experimental limits into constraints on neutrino mass, thus requires reliable knowledge of the NME. Accurate calculations of the $\Mzero$ are therefore essential for both the interpretation of current searches and the design of next-generation experiments~\cite{Agostini:2023}.
Despite decades of effort, theoretical uncertainties in $0\nu\beta\beta$ NMEs remain substantial, largely due to the model dependence of nuclear many-body approaches~\cite{Engel:2017,Yao:2022_PPNP,Agostini:2023}. These uncertainties are difficult to reduce within purely phenomenological frameworks, where each model relies on its own assumptions, effective interactions, and uncontrolled approximations.  

\textit{Ab initio} nuclear methods based on interactions derived from chiral effective field theory~\cite{Weinberg:1991,Epelbaum:2009RMP,Machleidt:2011PR} provide a systematically improvable framework for nuclear structure and decay~\cite{Lee:2009,Barrett:2013,Hagen:2014_RPP,Carlson:2015,Launey:2016,Stroberg:2019,Hergert:2020}. Within this framework, recent advances have enabled the first multi-method calculations of $0\nu\beta\beta$ NMEs in light- and medium-mass nuclei~\cite{Cirigliano:2019PRC,Basili:2020PRC,Yao:2021PRC}. As the lightest experimentally relevant $0\nu\beta\beta$ candidate and the isotope with the largest $Q$ value among known $2\nu\beta\beta$ emitters, $^{48}$Ca~\cite{Umehara:2008,NEMO-3:2016,CANDLES:2021} provides a unique benchmark for nuclear many-body methods~\cite{Senkov2013,Iwata:2016,Coraggio:2019,Gambacurta:2020}. Accordingly, the in-medium generator coordinate method (IM-GCM)~\cite{YJM2020}, the valence-space in-medium similarity renormalization group (VS-IMSRG)~\cite{Belley:2021PRL}, and coupled-cluster theory~\cite{Novario:2021} have all been applied to $^{48}$Ca using the same chiral nuclear interaction and decay operators, enabling direct cross-method comparisons beyond the reach of phenomenological models. These benchmark studies demonstrated that the resulting NMEs are consistent within the partially estimated theoretical uncertainties.  More recently, a rather full uncertainty quantification has been performed for the NME of $^{76}$Ge based on the IM-GCM and VS-IMSRG approaches~\cite{Belley:2024}. Moreover, the VS-IMSRG has been extended to most experimentally relevant candidate nuclei~\cite{Belley2023TeXe}. Nevertheless, uncertainties of approximately a factor of two remain, primarily due to truncations in the nuclear interactions and many-body model spaces. These truncations are systematically improvable in principle, but reducing them in practice poses substantial computational challenges~\cite{Belley:2024,Belley2023TeXe}.

To reduce the existing uncertainties, recent studies have explored correlations between $0\nu\beta\beta$ NMEs and experimentally accessible observables, including double Gamow--Teller (DGT) transitions~\cite{Shimizu:2018PRL,Yao:2022PRC,Belley:2022,Jokiniemi:2023PRC,Wang:2024PLB}, double-$\gamma$ transitions~\cite{Romeo:2021,Romeo:2025}, nucleon--nucleon scattering observables~\cite{Belley:2026_phase_shift}, high-energy probes of nuclear shape~\cite{Li:2025}, and low-energy nuclear structure properties~\cite{Horoi:2022,Belley:2022,Zhang:2024_short}. Among these observables, two-neutrino double-$\beta$ ($2\nu\beta\beta$) decay is of particular interest because its transition operator shares important similarities with that of $0\nu\beta\beta$ decay, while its half-life is being measured with increasingly high precision~\cite{NEMO-3:2016,Barabash2020,PandaX:2026}. Studies based on phenomenological nuclear models have suggested possible correlations between the corresponding matrix elements~\cite{Horoi:2022,Jokiniemi:2023PRC}. Whether such correlations persist in first-principles calculations, however, remains an open question. Addressing this question is nontrivial, as it requires an \textit{ab initio} framework capable of describing both decay modes consistently and reliably.

In this Letter, we develop a novel \textit{ab initio} in-medium no-core configuration-interaction (IM-NCCI) framework for general nuclear weak processes. In this approach, the in-medium similarity renormalization group (IMSRG) is employed to decouple the reference state of the initial nucleus from its excitations, while the NCCI method is used to describe the final states of the daughter nucleus. We demonstrate the predictive power of the framework through applications to GT transition strengths and to the NMEs governing the $2\nu\beta\beta$ and $0\nu\beta\beta$ decays of $^{48}$Ca. Applying the same framework to $0\nu\beta\beta$ decay, and including contributions from short-range operators, we obtain a total NME of $\Mzero=1.00\text{--}2.02$. Using 34 non-implausible chiral Hamiltonians, we further uncover a strong linear correlation between $\Mtwo$ and $\Mzero$. Combining this correlation with the experimental value of $\Mtwo$ yields a constrained prediction for the $0\nu\beta\beta$ NME, $\Mzero=1.30\text{--}1.65$. This work provides a first-principles foundation for constraining $0\nu\beta\beta$ NMEs in heavier candidate nuclei using experimentally accessible $2\nu\beta\beta$-decay data.

\paragraph{The IM-NCCI method.$-$}  We start from an intrinsic nuclear Hamiltonian composed of a kinetic term, two-body ($NN$) and three-body ($3N$) nuclear interaction terms,
\begin{equation}
\label{Eq:H}
H_0 = \sum_{i<j} \dfrac{(\mathbf{p}_i-\mathbf{p}_j)^2}{2M_NA} + \sum_{i<j} V_{ij}^{(NN)} +  \sum_{i<j<k} W_{ijk}^{(3N)},
\end{equation}
where $M_N$ is the nucleon mass, $A$ the mass number and $\mathbf{p}_i$ the momentum of the $i$-th nucleon.  Two different types of chiral interactions are used in this work, i.e., the EM1.8-2.0~\cite{Hebeler:2011,Entem:2003} and  $\Delta$N2LO$_{\rm GO}$(394)~\cite{Jiang:2020}. For the EM1.8-2.0 interaction, the $NN$ part is evolved via the SRG \cite{Bogner:2010} with resolution scale to $\lambda = 1.8$ fm$^{-1}$, while the $3N$ low-energy constants (LECs), $c_D$ and $c_E$, are fit to the \nuclide[3]{H} binding energy and the \nuclide[4]{He} matter radius~\cite{Hebeler:2011}. Besides, we adopt a set of 34 non-implausible samples~\cite{Hu:2022} in the 17-dimensional parameter space of the LECs around the $\Delta$N2LO$_{\rm GO}$(394) interactions \cite{Jiang:2020}.

The Hamiltonian $H_0$ in Eq.~(\ref{Eq:H}) is evolved through a sequence of unitary transformations to decouple a preselected reference state from its excitations within the IMSRG framework~\cite{Hergert:2016jk}. For the closed-shell nucleus $^{48}$Ca, we use a Hartree--Fock (HF) reference state $\ket{\Phi}$ and adopt the White generator to suppress off-diagonal couplings between the reference state and particle-hole excitations. For simplicity, we employ the normal-ordered two-body (NO2B) approximation, in which all operators are truncated at the normal-ordered two-body level, while contributions from higher-body operators are approximately incorporated through density-dependent lower-body terms. For $^{48}$Ca, this approximation is expected to be highly accurate. As a result, the reference state $\ket{\Phi}$ becomes an excellent approximation to the exact ground state of the evolved Hamiltonian~\cite{Tsukiyama11}. The final states of the daughter nuclei involved in the weak processes, such as \nuclide[48]{Sc} and \nuclide[48]{Ti}, are then obtained with the NCCI method using the evolved Hamiltonian. In this approach, the nuclear wave function is expanded in a basis of many-particle--many-hole ($mpmh$) configurations,
\be
\ket{\Psi}
= D_0\ket{\Phi}+
\sum_{mi} D^m_i \ket{\Phi^m_i}
+\sum_{mnij} D^{mn}_{ij} \ket{\Phi^{mn}_{ij}}
+\cdots ,
\label{eq:ciansatz}
\ee
where  the particle-hole excitaiton configurations are constructed as
\be
\ket{\Phi^m_i}=a^\dagger_m a_i\ket{\Phi}, \qquad
\ket{\Phi^{mn}_{ij}}=a^\dagger_m a^\dagger_n a_j a_i \ket{\Phi},
\ee
with analogous definitions for higher-order excitation configurations. The amplitudes $D$ are obtained with the sparse matrix diagonalization method. The main challenge of NCCI calculations is the exponential growth of the model-space dimension with increasing mass number $A$ and single-particle basis size.

To reduce the computational cost, we employ an importance-sampling algorithm~\cite{Roth:2009}, motivated by the observation that many basis states carry only very small amplitudes $D$ in the expansion of the low-lying-state wave functions in Eq.~(\ref{eq:ciansatz}). The importance truncation is implemented iteratively using a default threshold value of $\kappa_{\min}=1.0\times10^{-5}$. In addition, we calculate the NMEs for several values of $\kappa_{\min}$ and extrapolate to the limit $\kappa_{\min}\to0$ to estimate the uncertainty associated with the importance truncation \cite{Yao:2021PRC}. See Supplemental Material (SM) for more details.  Together with the conservation of the magnetic quantum number of angular-momentum $M$ and parity $\pi$, this leads to a block structure of the Hamiltonian matrix. With these considerations, we are able to extend  the model space for the $1^+_m$ states of odd-odd nucleus $^{48}$Sc up to $3p3h$ configurations, while treating the ground state ($0^+_1$) of $^{48}$Ti in a $4p4h$ space. 
For $0\nu\beta\beta$ decay, an alternative strategy is to decouple the reference state of the final nucleus, \nuclide[48]{Ti}. In this work, this reference state is constructed as a projected HFB state. Because model-space truncations are introduced, different choices of the reference state lead to different NME values. Following Refs.~\cite{Novario:2021,Belley:2021PRL}, we use this reference-state dependence to estimate the corresponding uncertainty in the predicted NMEs. 

\paragraph{The distribution of GT transition strength.$-$}  The GT$^-$ transition strengths from the ground state of \nuclide[48]{Ca} to the $1^+_m$ states of \nuclide[48]{Sc} have been measured through the \nuclide[48]{Ca}$(p,n)$ reaction~\cite{Yako:2009}, providing a stringent benchmark for the IM-NCCI method. The corresponding GT transition strength is computed as
\be
B(\GT^-; 0^+_1\rightarrow 1^+_m)
=\left|
\bra{1^+_m} \boldsymbol{\sigma}\tau^- \ket{0^+_1}
\right|^2 .
\label{eq:bgt}
\ee

\begin{figure}[bt]
\safeincludegraphics[width=0.8\columnwidth]{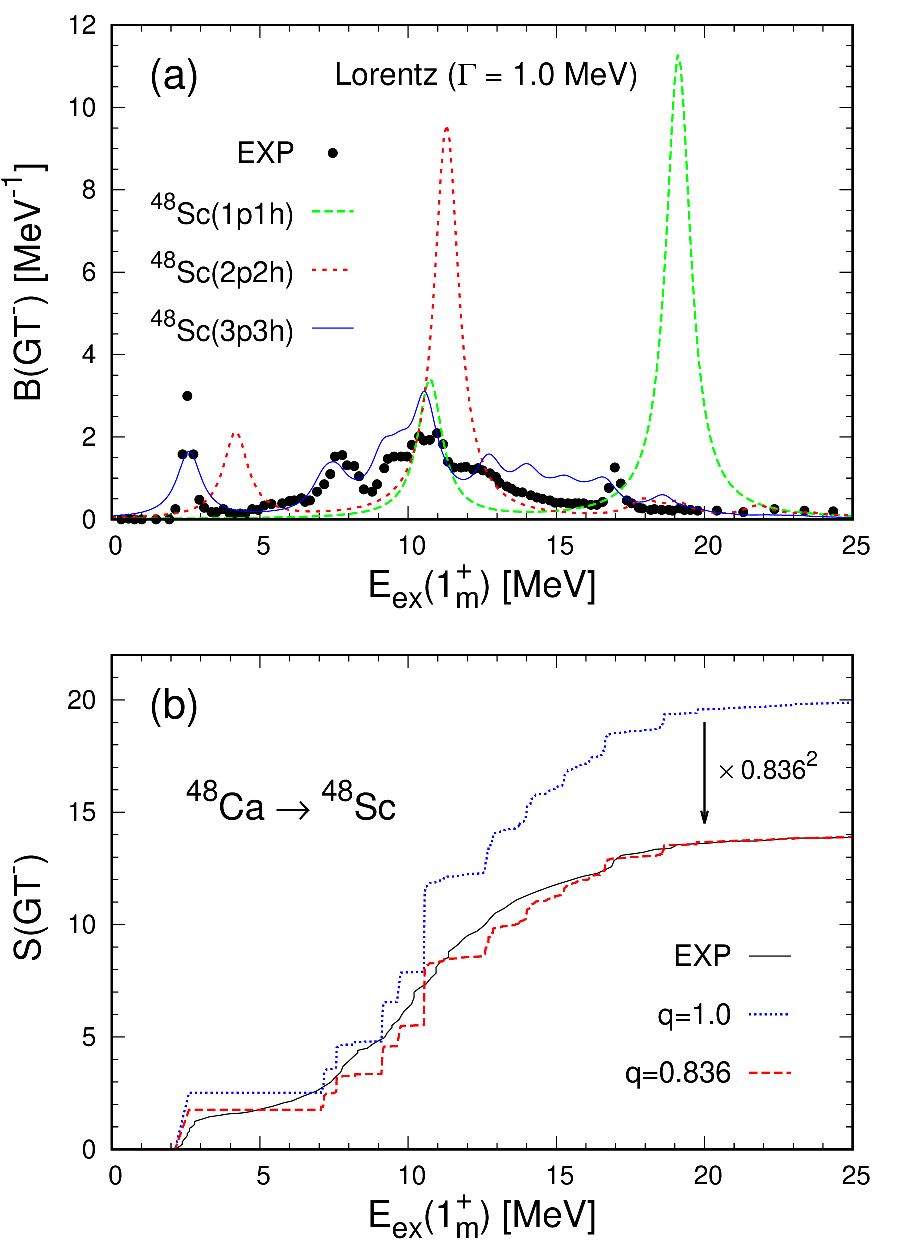}
  \caption{(Color online) (a) Distribution of the GT$^-$ transition strength, $B({\rm GT}^-;0^+_1\to 1^+_m)$, from $^{48}$Ca to the $1^+_m$ states of $^{48}$Sc as a function of excitation energy $E_{\rm ex}(1^+_m)$. The IM-NCCI results are shown without any quenching factor and are folded with a Lorentzian of width $\Gamma=1$ MeV. Experimental data from Ref.~\cite{Yako:2009} are shown for comparison.
(b) Cumulative GT$^-$ strength for $^{48}$Ca$\rightarrow{}^{48}$Sc, compared with experiment. The unquenched result (blue) and the result scaled by the quenching factor $q=0.836$ (red) are both shown.}
  \label{fig:gtminus}
\end{figure}

Figure~\ref{fig:gtminus} shows the results obtained with the chiral interaction EM1.8-2.0 as a function of the excitation energy of the $1^+_m$ states, compared with the experimental GT$^-$ distribution extracted from the \nuclide[48]{Ca}$(p,n)$ data of Ref.~\cite{Yako:2009}. The calculation including up to 3p3h configurations reasonably reproduces the positions of several major peaks, including those around $E_{\rm ex}\simeq 3$, $8$, and $11$ MeV. Reproducing these features has been challenging for many phenomenological nuclear models~\cite{Niu:2014PRC,Gambacurta:2020,Yang:2022}, yet it is essential for a reliable description of $2\nu\beta\beta$ decay. A substantial improvement is observed when the model space for \nuclide[48]{Sc} is enlarged up to $3p3h$ configurations. On the other hand, the strength in the excitation-energy region $E_{\rm ex}\simeq 12$--$16$ MeV is somewhat overestimated. Consequently, the cumulative GT$^-$ strength,
\be
S_{\GT^-}
=\sum_{m}
B(\GT^-;0^+_1\rightarrow 1^+_m),
\ee
is clearly overestimated for $E_{\rm ex}>12$ MeV compared with experiment, as shown in Fig.~\ref{fig:gtminus}. The Ikeda sum rule,
\be
S_{\GT^-}-S_{\GT^+}=3(N-Z),
\label{eq:ikeda}
\ee
provides a model-independent check on the completeness of the calculated states. For a neutron-rich nucleus such as \nuclide[48]{Ca}, one has $S_{\GT^-} \gg S_{\GT^+}$, so the sum rule is dominated by the total GT$^-$ strength. The theoretical integrated strength below 25 MeV is about 20, compared with the experimental value of about 14. The ratio between the experimental and theoretical strengths is therefore $q^2=0.836^2$, which is interpreted as a quenching factor associated with missing two-body-currents (2BCs), as demonstrated in previous \textit{ab initio} studies~\cite{Gysbers:2019df,Novario:2021,Li:2026}.

\paragraph{The NME of $2\nu\beta\beta$ decay.$-$} We then apply the IM-NCCI method to $2\nu\beta\beta$ decay in \nuclide[48]{Ca}, which has been studied extensively with various nuclear models~\cite{Horoi:2007,Iwata:2015,Simkovic:2018,Kostensalo:2020,Novario:2021,Miskiewicz:2025,Chen:2026}. Reproducing $\Mtwo$ without introducing free parameters, however, remains a significant challenge~\cite{Yako:2009,Barabash2020,Miskiewicz:2025}. The inverse half-life of $2\nu\beta\beta$ decay is written as
\be
[T^{2\nu}_{1/2}]^{-1}
=
G_{2\nu} (qg_A)^4 \left|m_e\Mtwo\right|^2,
\label{eq:2nuhalflife}
\ee
where $G_{2\nu}\simeq 1.6 \times 10^{-17}~{\rm yr}^{-1}$ is the phase-space factor~\cite{Kotila:2012,Simkovic:2018}. The symbol $q$ denotes the quenching factor applied to $g_A$. The electron mass, $m_e=0.511$ MeV, is included so that the combination $m_e \Mtwo$ is dimensionless. The NME $\Mtwo$ is calculated as~\cite{Suhonen:1998,Avignone:2008,Chen:2026}
\be
\Mtwo=
\sum_m
\frac{
\ME{0_f^+}{\sigma\tau^-}{1_m^+}
\ME{1_m^+}{\sigma\tau^-}{0_i^+}
}{
E_m - (E_i + E_f)/2
},
\label{eq:m2nu}
\ee
where the sum runs over all intermediate $1^+_m$ states in \nuclide[48]{Sc}. The numerator is the product of the GT$^-$ transition matrix element from \nuclide[48]{Ca} to \nuclide[48]{Sc} and the corresponding GT$^+$ matrix element connecting \nuclide[48]{Ti} to \nuclide[48]{Sc}. In this work, the same set of intermediate $1^+_m$ states is used consistently for both GT branches. This treatment differs from the conventional QRPA approach to $\Mtwo$, in which the intermediate states reached from the initial and final nuclei are usually generated separately~\cite{Simkovic:2007,Fang:2011,Mustonen:2013,Lv:2021,Yao:2022_PPNP}.

\begin{figure}[bt]
  \safeincludegraphics[width=\columnwidth]{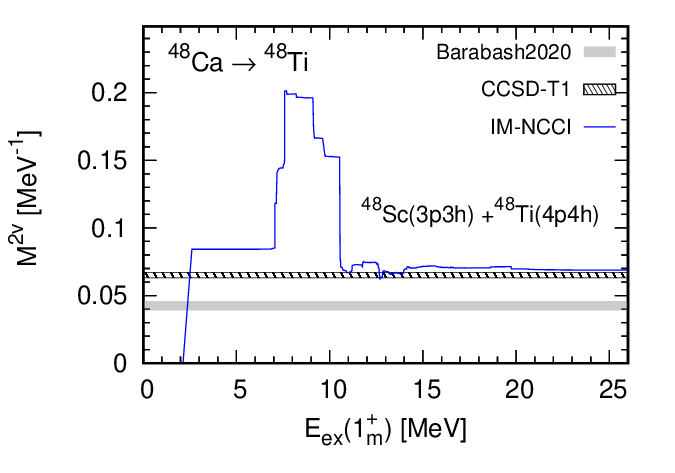}
  \caption{(Color online) Running sum of the $\Mtwo$ for
  $^{48}$Ca$\rightarrow{}^{48}$Ti as a function of the excitation energy of the $1^+_m$ state in $^{48}$Sc. The results are compared with the value from the CCSD-T1 calculation~\cite{Novario:2021} and experimental data~\cite{Barabash2020}.}
  \label{fig:m2nu}
\end{figure} 
 
Figure~\ref{fig:m2nu} shows the running sum of $\Mtwo$ as a function of the excitation energy of the intermediate $1^+_m$ states. The low-lying $1^+_m$ states provide the dominant contribution, whereas contributions from higher-energy states largely cancel each other. This behavior is expected for $^{48}$Ca and illustrates how the low-energy GT transitions dominate the total NME of $2\nu\beta\beta$ in \nuclide[48]{Ca}.
Our IM-NCCI calculation gives $M^{2\nu}_{\rm LO}=0.066$--$0.077~{\rm MeV}^{-1}$, which is consistent with the CCSD-T1 result, $M^{2\nu}_{\rm LO}=0.065(2)~{\rm MeV}^{-1}$~\cite{Novario:2021}.


After applying the quenching factor $q=0.836$ extracted from the GT sum, which accounts for missing higher-order 2BCs, we obtain $M^{2\nu}=0.046$--$0.054~{\rm MeV}^{-1}$ from the IM-NCCI calculation. This value is close to the experimental result, $M^{2\nu}=0.0425(3)~{\rm MeV}^{-1}$~\cite{NEMO-3:2016,Barabash2020}, and with the quenched CCSD-T1 result, $0.042(1)~{\rm MeV}^{-1}$~\cite{Novario:2021}. This agreement provides an important validation of the IM-NCCI framework, since $2\nu\beta\beta$ decay is highly sensitive to the detailed spectroscopy of the intermediate odd-odd nucleus and therefore represents a stringent benchmark.

\paragraph{The NME of $0\nu\beta\beta$ decay.$-$} 
Assuming the standard light-Majorana-neutrino exchange mechanism, the inverse half-life of $0\nu\beta\beta$ decay is given by
\begin{equation}
\label{half-life}
 [T^{0\nu}_{1/2}]^{-1} = g^4_A G_{0\nu}
 \left\lvert\dfrac{\langle m_{\beta\beta}\rangle}{m_e}\right\rvert^2
 \left\lvert M^{0\nu}\right\rvert^2 ,
\end{equation}
where $m_e$ is the electron mass, $g_A$ is the axial-vector coupling constant, and
$G_{0\nu}\simeq 2.6\times 10^{-14}\,{\rm yr}^{-1}$ is the phase-space factor
\cite{Kotila:2012,G_value,Simkovic:2018}. The $\Mzero$ is evaluated within the closure approximation as
\begin{equation}
\Mzero = \Mzero_{\rm LR}+\Mzero_{\rm SR}
= \bra{0^+_f} \sum_\alpha \hat O^{0\nu}_\alpha \ket{0^+_i} .
\label{eq:m0nudecomp}
\end{equation}
Here, the index $\alpha$ labels the components of the transition operator $\hat O^{0\nu}_\alpha$, including the three long-range (LR) terms, Gamow--Teller (GT), Fermi (F), and tensor (T) contributions~\cite{Simkovic99,Yao:2022PRC}, as well as the short-range (SR) contact term~\cite{Cirigliano:2018,Wirth:2021}.

Following previous \textit{ab initio} studies~\cite{YJM2020,Belley:2021PRL,Novario:2021}, we employ the same interaction, EM1.8/2.0, and transition operator to compute the NME $\Mzero$ for \nuclide[48]{Ca}, extrapolating the threshold value $\kappa_{\min}$ to zero; see the SM for details. The long-range contribution is found to be $M^{0\nu}_{\rm LR}=0.60$--$1.14$. The contact contribution is estimated as $M^{0\nu}_{\rm SR}=0.40$--$0.88$. Here, the uncertainty arises from the reference-state dependence, the finite basis size $\eMax$, and the choice of the LEC $g_\nu^{NN}$ associated with the contact transition operator~\cite{Wirth:2021}. The resulting total NME is therefore $M^{0\nu}=1.00\text{--}2.02$. As shown in SM, considering the uncertainty,  the value $M^{0\nu}_{\rm LR}$ of the long-range transition operators is consistent with the results of the previous \textit{ab initio} calculations. Together with the experimental lower limit on the half-life, $T^{0\nu}_{1/2}\geq 5.8\times10^{22}$ yr~\cite{Umehara:2008}, this NME range implies an upper limit on the effective Majorana neutrino mass of $\langle m_{\beta\beta}\rangle \lesssim 4.1\text{--}8.6~\mathrm{eV}$. Since the same chiral nuclear interaction and transition operator are used in four independent \textit{ab initio} calculations, namely IM-GCM, VS-IMSRG, CCSD-T1, and IM-NCCI (see the SM), the spread of the resulting NMEs can be interpreted as an estimate of the uncertainty associated with the different many-body truncation schemes employed in these methods. Additional sources of uncertainty, however, must be considered to obtain the final NME uncertainty~\cite{Belley:2024}, including the truncation of chiral nuclear forces and the omission of higher-order two-body currents. The final NME uncertainty is therefore expected to be larger than the present estimate.

\begin{figure}[tb]
  \safeincludegraphics[width=\columnwidth]{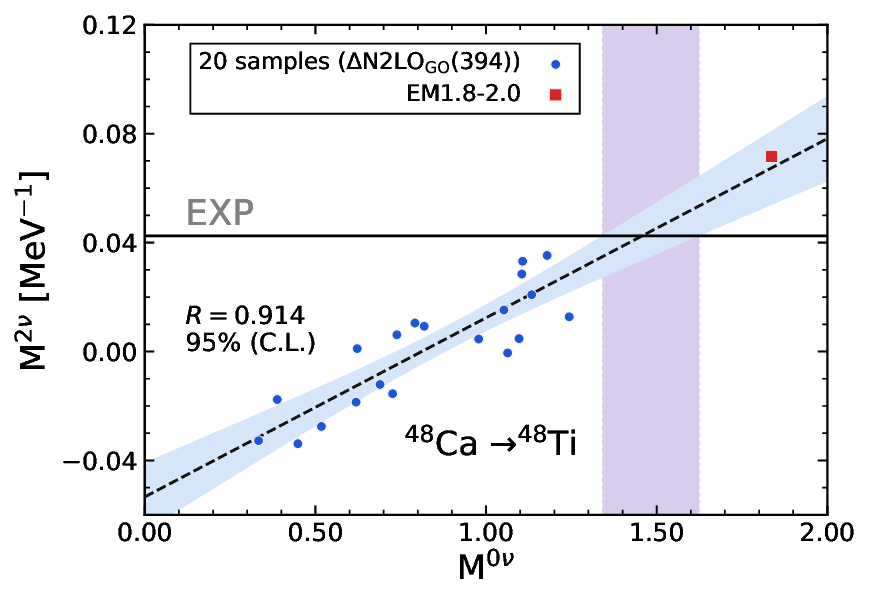}
  \caption{(Color online) Correlation between the NMEs of $M^{0\nu}$ and $M^{2\nu}$ for \nuclide[48]{Ca} $\rightarrow$ \nuclide[48]{Ti}.  The blue circles denote results obtained from 20 samples of chiral Hamiltonians, while the red square corresponds to the result of EM1.8-2.0 interaction.  The dashed line shows the linear regression, and the shaded band indicates the  95\% confidence interval for the correlation relation. } 
  \label{fig:Correlation_0nu_2nu}
\end{figure}

\paragraph{Exploring correlations with  $2\nu\beta\beta$ decay and DGT transition.$-$} Finally, we explore the correlation between the NMEs $\Mzero$ and $\Mtwo$ within the IM-NCCI framework, which may provide a pathway toward reducing the remaining uncertainties in \textit{ab initio} predictions of the $0\nu\beta\beta$ NME. To this end, we calculate the NMEs $\Mzero$ and $\Mtwo$ using a set of chiral Hamiltonians, including 34 samples of LECs based on the $\Delta$N2LO$_{\rm GO}$(394) interaction~\cite{Jiang:2020,Hu:2022}. Some samples give a poor description of the lowest $1^+$ state in \nuclide[48]{Sc}. Therefore, we apply a quality filter and retain only those satisfying 
 $E_{\rm ex}^{\rm cal}(1_1^+) \in
[0.3,1.7]\,E_{\rm ex}^{\rm exp}(1_1^+)$. This leaves 20 samples. The resulting correlation between $\Mzero$ and $\Mtwo$ is shown in Figure~\ref{fig:Correlation_0nu_2nu}, while the corresponding result for all 34 samples is presented in the SM. We note that the effects of 2BCs on the NMEs are not included. Since no ab initio study has yet quantified these effects for both $\Mzero$ and $\Mtwo$, the magnitude of the corresponding quenching remains an open question. As discussed for GT transitions, the 2BCs may reduce $\Mtwo$ by about 30\%, comparable to the quenching of $\Mzero$ found in phenomenological nuclear models~\cite{Menendez:2011,Jokiniemi:2023PRC}. Although a more careful treatment reported a much smaller effect~\cite{Wang2018}, possible additional contributions from divergent terms require further investigation. For this reason, we present the correlation between $\Mzero$ and $\Mtwo$ without including the 2BCs effects for both NMEs.
It is shown in Fig.~\ref{fig:Correlation_0nu_2nu} that although both $\Mzero$ and $\Mtwo$ exhibit a broad spread, reflecting their sensitivity to the underlying LECs of chiral nuclear interactions, the two NMEs display a strong linear correlation. The Pearson coefficient increases from $R=0.82$ for the full set to $R=0.91$ after applying the filter. In both cases, these values are in remarkable agreement with those obtained in the valence-space shell-model study based on randomly generated nuclear interaction matrix elements around the optimal values~\cite{Horoi:2022}. They are also consistent with pnQRPA and shell-model studies employing fixed effective nuclear forces~\cite{Jokiniemi:2023PRC}. In the latter case, however, the correlations were established across different isotopes and therefore involve an additional dependence on the mass number $A$.

Moreover, using the same set of 34 chiral interactions, we re-examine the correlation between the $0\nu\beta\beta$ NME and the double-GT transition matrix element in \nuclide[48]{Ca}. We find a strong correlation, with a Pearson coefficient of $R\simeq0.92$; see the SM. This correlation is even stronger than the value $R=0.84$ obtained in the VS-IMSRG study of \nuclide[76]{Ge} using the same 34 sets of chiral interactions~\cite{Belley:2022}. This result is not in contradiction with the earlier \textit{ab initio} study of Ref.~\cite{Yao:2022PRC}, where only a weak correlation between double-GT and $0\nu\beta\beta$ NMEs was found for isospin-changing processes across different isotopes. In that case, the correlation was also influenced by an additional dependence on $A$.

In short, our results demonstrate strong linear correlations among the NMEs governing $0\nu\beta\beta$ and $2\nu\beta\beta$ decays and DGT transitions in \nuclide[48]{Ca} from ab initio calculations. These correlations provide a first-principles foundation for using measured $2\nu\beta\beta$ and DGT NMEs to constrain the $0\nu\beta\beta$ NME in experimentally relevant candidate nuclei. At present, however, no experimental data are available for the DGT NME. Using the available $2\nu\beta\beta$-decay data together with our correlation relation at the 95\% confidence level, we constrain the total NME to $\Mzero=1.30\text{--}1.65$, as shown in Figure~\ref{fig:Correlation_0nu_2nu}. This range is consistent with our directly derived estimate, but has a significantly reduced uncertainty.

\paragraph{Summary.$-$}We have developed a novel \textit{ab initio} in-medium no-core configuration-interaction (IM-NCCI) framework based on chiral NN+3N interactions and applied it to Gamow--Teller (GT) transition strengths and to the nuclear matrix elements (NMEs) governing $2\nu\beta\beta$
and $0\nu\beta\beta$ decays in $^{48}$Ca. This framework enables, for the first time from a first-principles perspective, a systematic investigation of the correlation between the
NMEs of $2\nu\beta\beta$ and $0\nu\beta\beta$ decays. The locations of several main resonance peaks in the Gamow–Teller (GT) strength distribution for  \nuclide[48]{Ca} $\to$ \nuclide[48]{Ti} are reasonably reproduced. The cumulative GT strength indicates missing contributions from two-body currents (2BCs),  which corresponds to a quenching factor of $q=0.836$.
Incorporating this factor yields a $2\nu\beta\beta$ NME in excellent agreement with experiment. 

Applying the same framework to $0\nu\beta\beta$ decay, we obtain long-range NMEs consistent with previous \textit{ab initio} calculations. After including the contribution from short-range
transition operators, we predict a total NME, 
$M^{0\nu}=1.00\text{--}2.02$. The quoted uncertainty reflects only the model-space truncation of the IM-NCCI framework and the uncertainty in the low-energy constant $g_\nu^{NN}$.  Using 34 non-implausible chiral Hamiltonians, we uncover strong linear correlations between $\Mzero$ and the NMEs governing $2\nu\beta\beta$ and double GT decays. Combining these correlations with the experimental $2\nu\beta\beta$-decay data yields a constrained prediction
for the $0\nu\beta\beta$ NME, $\Mzero = 1.30\text{--}1.65$, with substantially reduced uncertainty.

This work establishes IM-NCCI as a complementary \textit{ab initio} many-body framework for describing general nuclear weak processes. It also provides a foundation for constraining $0\nu\beta\beta$ NMEs in heavier candidate nuclei using experimentally accessible information from $2\nu\beta\beta$ decay and double GT transitions. Future work will focus on a quantitative treatment of 2BCs contributions to both $2\nu\beta\beta$ and $0\nu\beta\beta$ decays, as well as to other weak processes, including superallowed Fermi transitions, for which sub-percent theoretical precision is required for meaningful tests of CKM unitarity~\cite{Hardy:2020}.

\paragraph{Acknowledgments.} 
We thank H. Ejiri, J. Engel, K. Hagino, H. Hergert, G. Li, Y. F. Niu, and K. Yoshida for helpful discussions, and A. Belley, W. L. Lv, and Y. K. Wang for their careful reading of the manuscript and valuable suggestions. This work was supported in part by the National Natural Science Foundation of China under Grant Nos. 125B2108, 12575126, 12405143, 12375119, and 12141501.

X. Lian and C. R. Ding contributed equally to this work.

 
%



\section*{Supplemental material for:  \textit{Ab initio} correlations between neutrinoless and two-neutrino double-beta decays in $^{48}$Ca}


\subsection{The importance sampling algorithm in the IM-NCCI method}
The dimension of configurations in the IM-NCCI method grows exponentially with the nucleon number and single-particle basis size. To mitigate the computation demanding, we adopt the importance-truncation strategy~\cite{Roth:2009}. Specifically, starting from a reference space ${\cal M}_{\mathrm{ref}}$ and reference eigenstates $\ket{\Psi_k^{\mathrm{ref}}}$,
\beq
H_{\mathrm{ref}}\ket{\Psi_k^{\mathrm{ref}}}=E_k^{(0)}\ket{\Psi_k^{\mathrm{ref}}},
\eeq
we define the zeroth-order Hamiltonian
\beq
H_0=\sum_{k\in {\cal M}_{\mathrm{ref}}}E_k^{(0)}
\ket{\Psi_k^{\mathrm{ref}}}\bra{\Psi_k^{\mathrm{ref}}}
+\sum_{v\notin {\cal M}_{\mathrm{ref}}}\epsilon_v \ket{\Phi_v}\bra{\Phi_v},
\eeq
with $\epsilon_v=\bra{\Phi_v}H\ket{\Phi_v}$. Treating $H-H_0$ perturbatively leads to the importance measure
\beq
\kappa_v = \sum_{k\in {\cal M}_{\mathrm{ref}}}
\left|
\frac{\bra{\Phi_v}H\ket{\Psi_k^{\mathrm{ref}}}}
{\epsilon_v-E_k^{(0)}}
\right| .
\label{eq:kappa}
\eeq
Configurations with $\kappa_v \ge \kappa_{\min}$ are retained in the truncated space,
\beq
{\cal M}(\kappa_{\min}) = {\cal M}_{\mathrm{ref}} \oplus
\mathrm{span}\{\ket{\Phi_v}\,|\,\kappa_v\ge \kappa_{\min}\}.
\eeq
After diagonalization in ${\cal M}(\kappa_{\min})$, the procedure can be iterated by promoting the resulting truncated space to the next reference space. In this way, the dominant physics of much larger spaces is retained while significantly reducing the dimensionality.

\begin{figure}[bt]
  \safeincludegraphics[width=0.8\columnwidth]{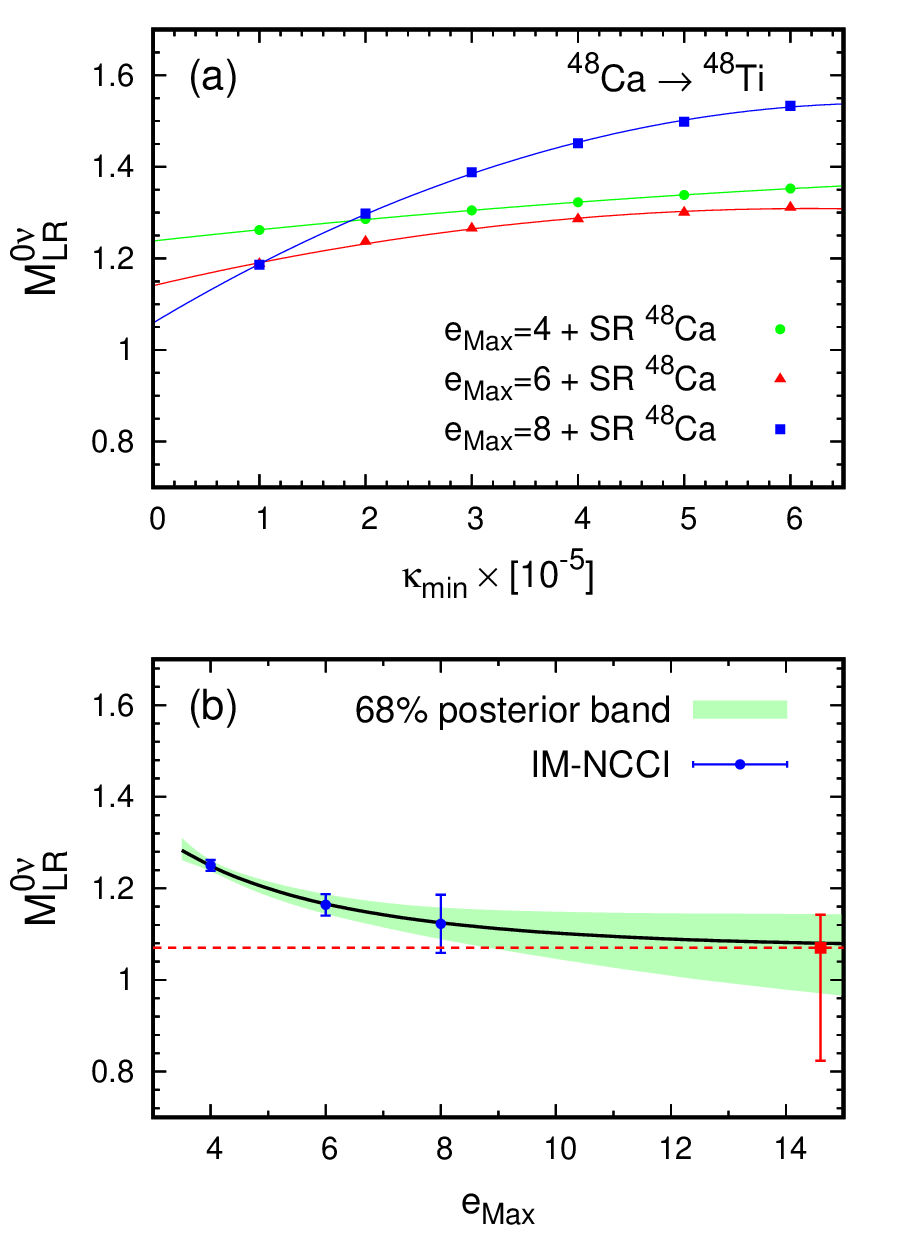}
  \caption{Threshold and model-space extrapolation of the long-range part of the $0\nu\beta\beta$ matrix element, $M^{0\nu}_{\rm LR}$, for $^{48}$Ca$\rightarrow{}^{48}$Ti using the $^{48}$Ca single-reference (SR) state. Panel (a) shows the dependence on the importance cutoff $\kappa_{\min}$ for $\eMax=4,6,8$; the curves indicate extrapolations toward $\kappa_{\min}=0$. Panel (b) shows the extrapolated values as a function of $\eMax$, together with the 68\% posterior band used to estimate the residual $\eMax$-truncation uncertainty.}
  \label{fig:kappa_N}
\end{figure}

\begin{figure}[hbt]
  \safeincludegraphics[width=0.7\columnwidth]{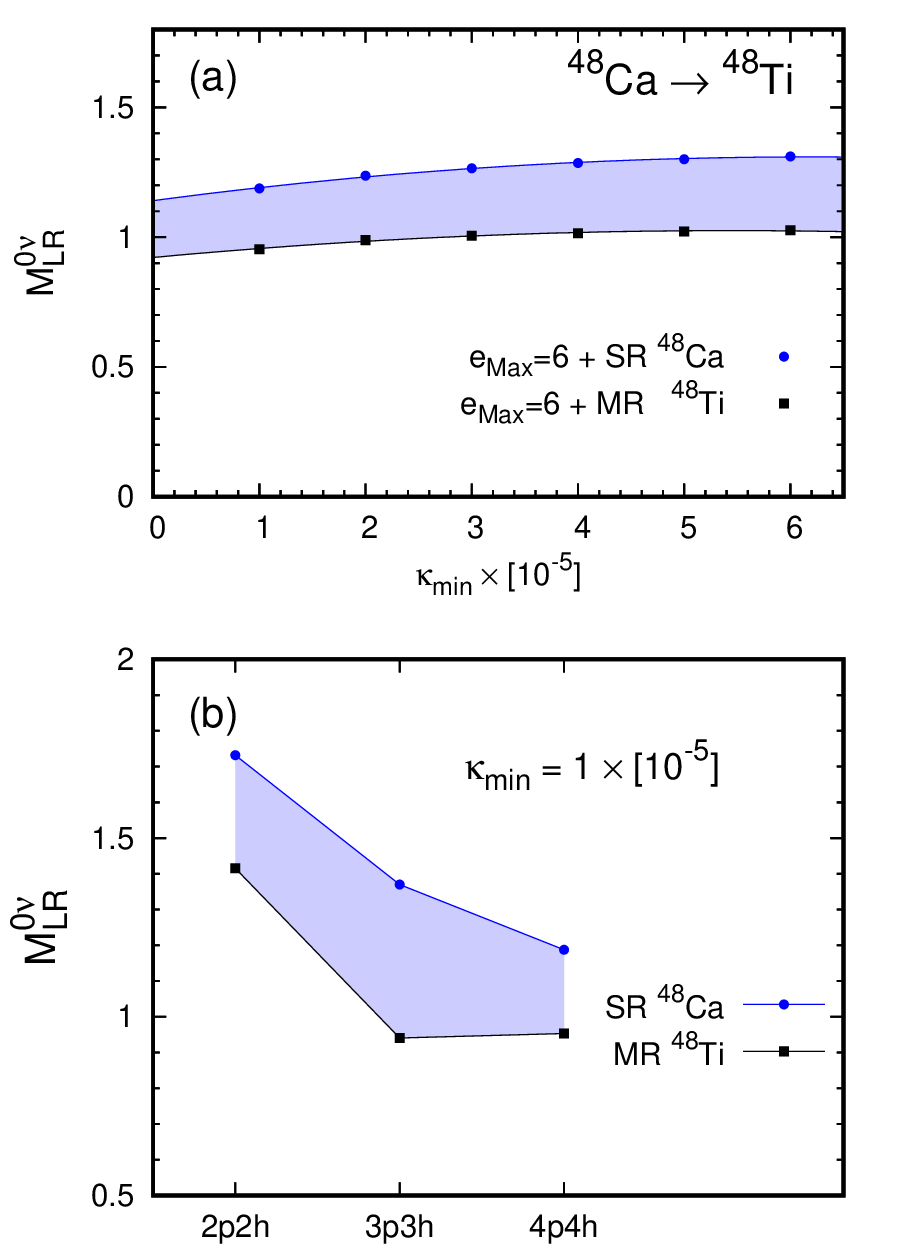}
  \caption{Reference-state and excitation-rank dependence of $M^{0\nu}_{\rm LR}$ for $^{48}$Ca$\rightarrow{}^{48}$Ti. Panel (a) compares the $\kappa_{\min}$ extrapolations at $\eMax=6$ obtained from a $^{48}$Ca single-reference (SR) state and a $^{48}$Ti multi-reference (MR) state; the shaded band indicates the range spanned by the zero-threshold extrapolations. Panel (b) shows the same two reference choices at the default cutoff $\kappa_{\min}=1.0\times10^{-5}$ as the particle-hole excitation rank is increased from $2p2h$ to $4p4h$.}
  \label{fig:kappa_ref}
\end{figure}

\begin{figure}[bt]
  \safeincludegraphics[width=\columnwidth]{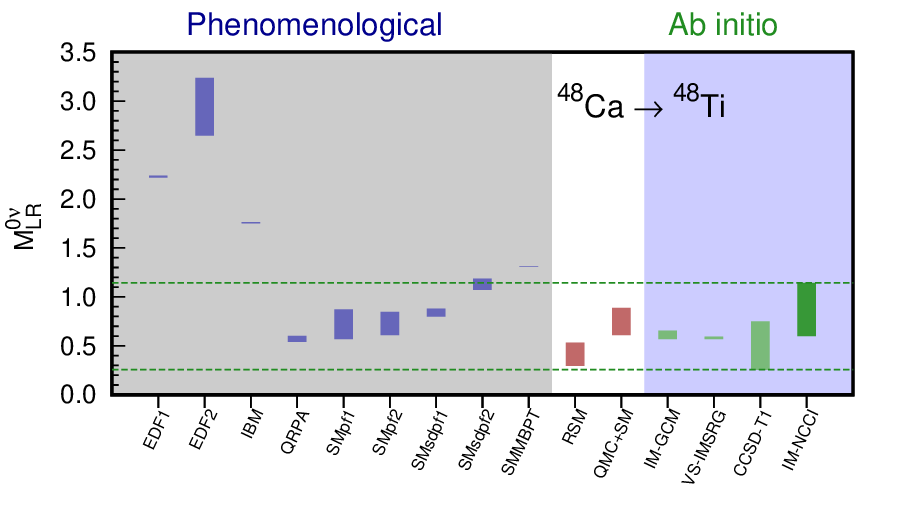}
  \caption{(Color online) The NMEs of $0\nu\beta\beta$ decay in \nuclide[48]{Ca} obtained in the present IM-NCCI calculation, compared with previous results from phenomenological~\cite{Vaquero:2013,Yao:2015,Barea:2015,Simkovic:2013,Menendez:2009,Senkov2013,Iwata:2016,Kwiatkowski:2014,Coraggio:2020,Weiss:2022} and \textit{ab initio}~\cite{YJM2020,Belley:2021PRL,Novario:2021} nuclear many-body approaches, where only the value by the long-range transition operators is considered.  The upper and lower dashed lines correspond to  1.14 and 0.26, respectively.}
  \label{fig:compare_NME40nu}
\end{figure}

\begin{figure}[bt]
  \safeincludegraphics[width=0.7\columnwidth]{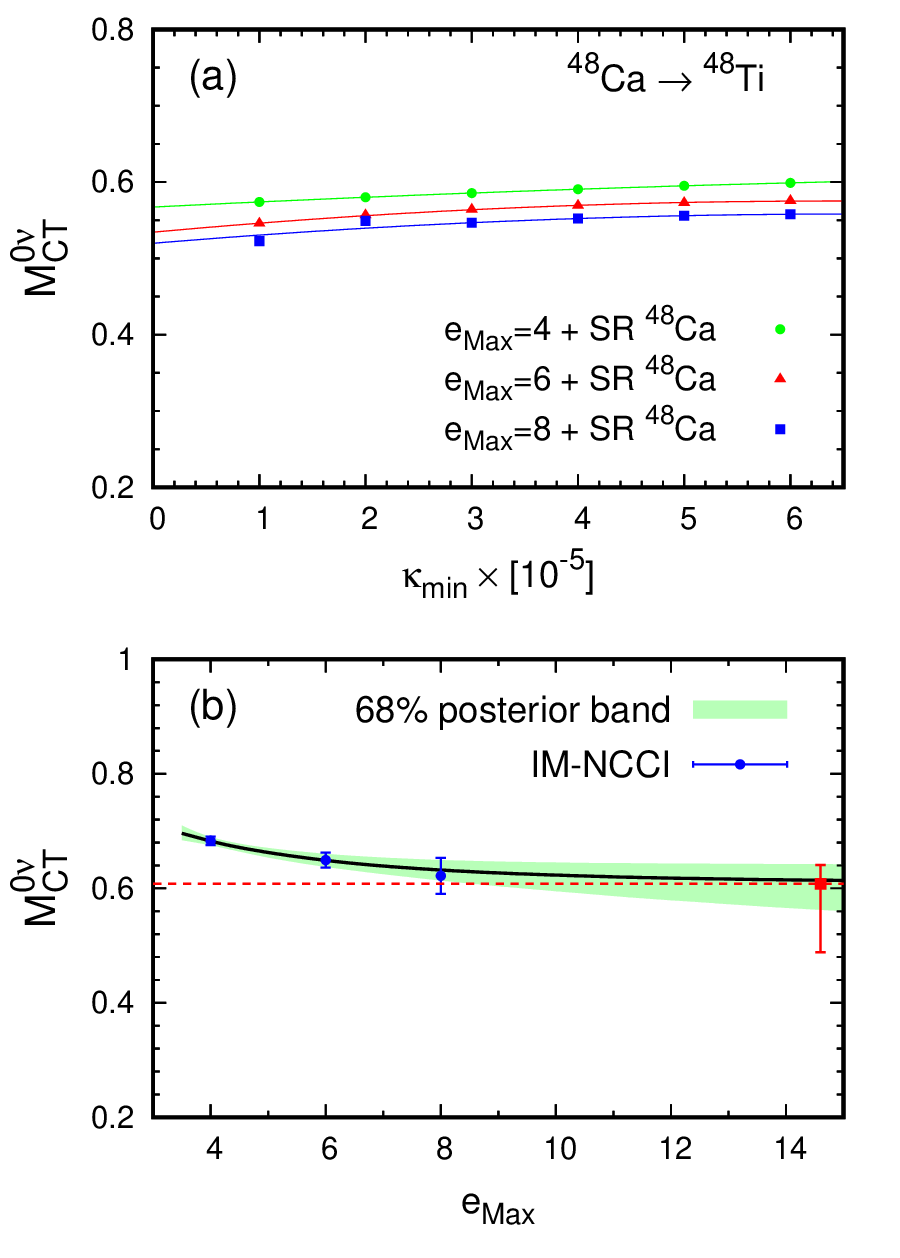}
  \caption{The same as Figure~\ref{fig:kappa_N}, but for the NME $M^{0\nu}_{\rm SR}$ of the short-range contact transition operator.}
  \label{fig:CT_kappa_eMax}
\end{figure}

\begin{figure}[bt]
   \safeincludegraphics[width=0.7\columnwidth]{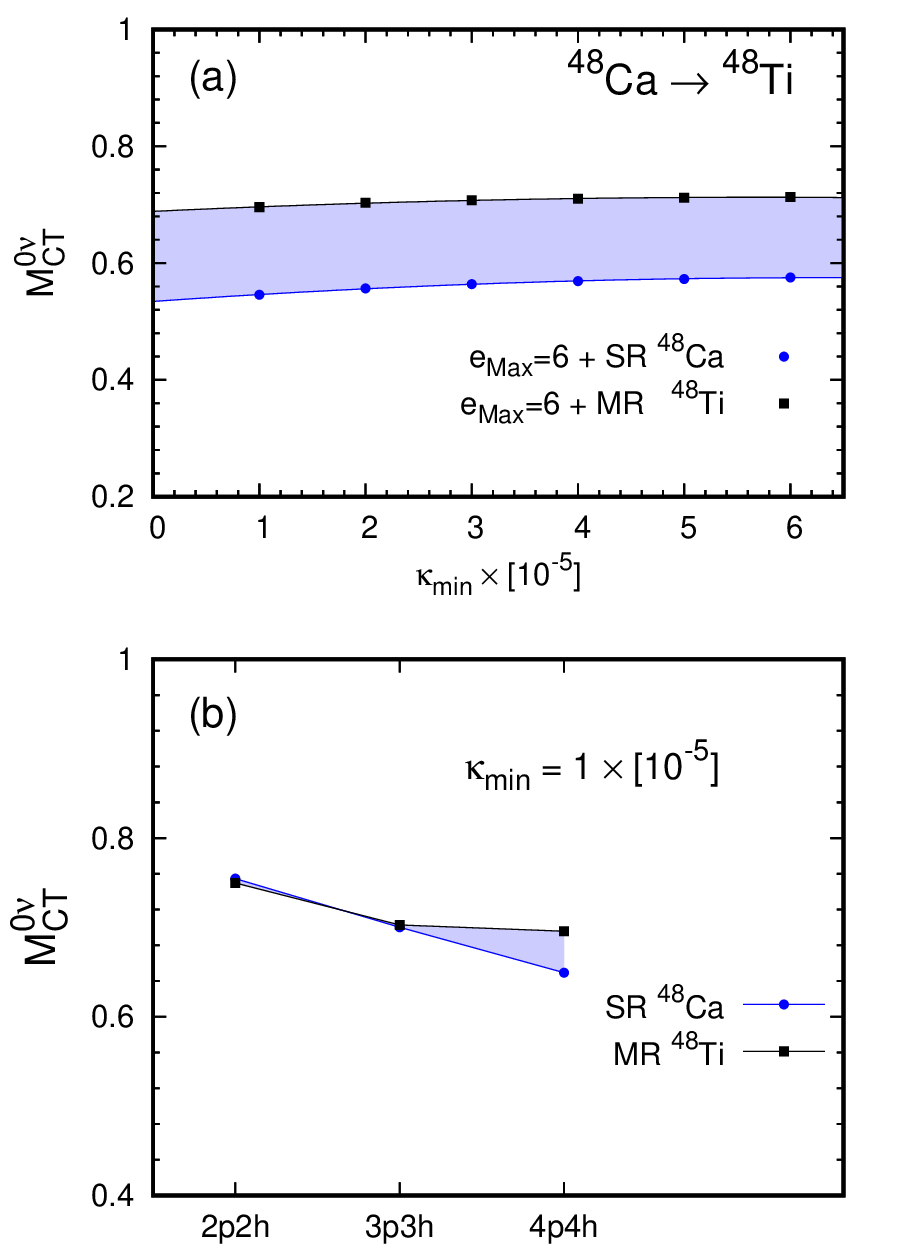}
 \caption{The same as Figure~\ref{fig:kappa_ref}, but for the NME $M^{0\nu}_{\rm SR}$ of the short-range contact transition operator.}
  \label{fig:CT_ref}
\end{figure}

\begin{figure}[bt]
   \safeincludegraphics[width=0.8\columnwidth]{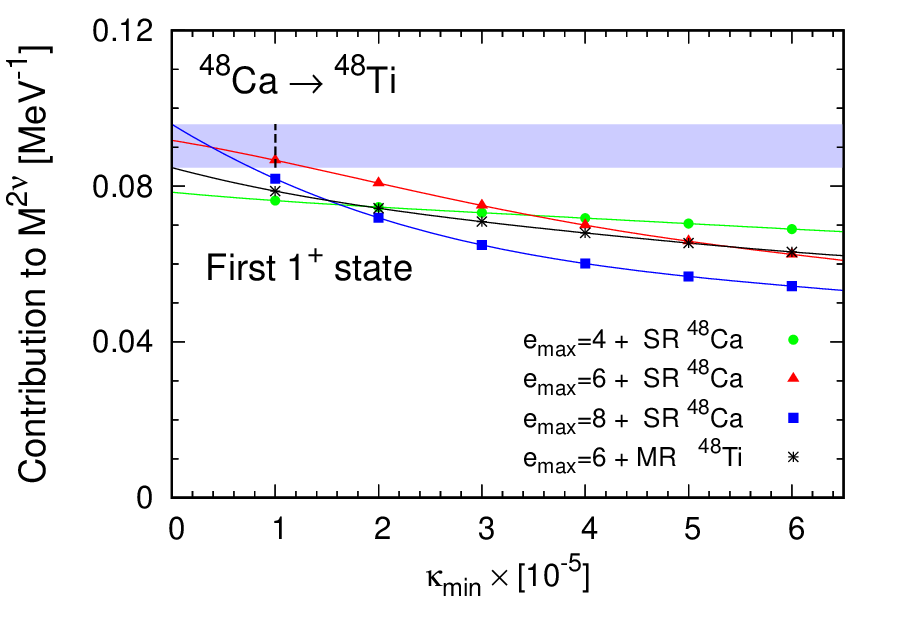}
  \caption{
  Threshold dependence of the contribution of the lowest $1^+$ intermediate state in $^{48}$Sc to the $M^{2\nu}$ for the decay $^{48}$Ca$\rightarrow{}^{48}$Ti. Results are shown for $^{48}$Ca SR reference states with $\eMax=4, 6, 8$ and for a $^{48}$Ti MR reference state with $\eMax=6$. The vertical dashed line denotes the default importance cutoff, $\kappa_{\min}=1.0\times10^{-5}$. The shaded band indicates the uncertainty estimate obtained from the threshold extrapolations of this dominant intermediate-state contribution. The black dashed lines show the corresponding uncertainty interval after shifting this band to the total matrix element calculated at $\eMax=6$.
  }
  \label{fig:kappa_2v}
\end{figure}

\begin{figure}[bt]
   \safeincludegraphics[width=0.8\columnwidth]{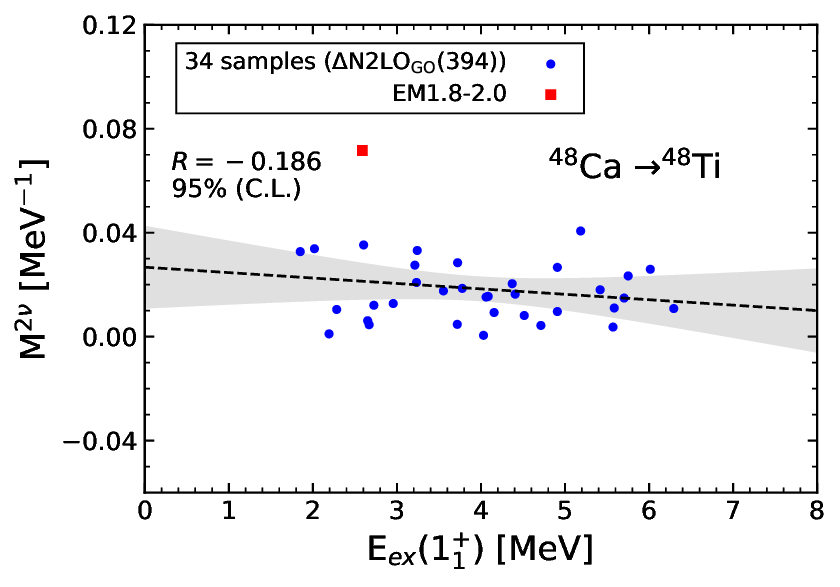}
  \caption{The $M^{2\nu}$ as a function of the excitation energy of the lowest $1^+$ state in $^{48}$Sc from the IM-NCCI calculation using all the 34 samples. Blue circles denote the 34 sets of sampled interactions, while the red square marks the EM1.8--2.0 result. The dashed line represents the linear regression, and the gray band indicates the 95\% confidence interval. }
  \label{fig:N_extra}
\end{figure}

\subsection{The convergence behavior of NMEs and uncertainty estimation}
Here we focus on the uncertainties in the NMEs arising from the importance cutoff $\kappa_{\min}$, the finite single-particle basis size $\eMax$, the choice of reference state, and the particle-hole excitation-rank truncation.

Figures~\ref{fig:kappa_N}(a) and \ref{fig:kappa_N}(b) show the long-range NME $M^{0\nu}_{\rm LR}$ as a function of the threshold $\kappa_{\min}$ and the basis size $\eMax$, respectively. Following Ref.~\cite{Yao:2021PRC}, we extrapolate the NME to the limits $\kappa_{\min}\to0$ and $\eMax\to\infty$, obtaining
$M^{0\nu}_{\rm LR}=0.82\text{--}1.14$.

We next estimate the uncertainties associated with the choice of reference state and the particle-hole excitation-rank truncation. Figures~\ref{fig:kappa_ref}(a) and \ref{fig:kappa_ref}(b) show, respectively, the reference-state dependence of $M^{0\nu}_{\rm LR}$ and its convergence with respect to the particle-hole excitation rank. As shown in Fig.~\ref{fig:kappa_ref}(a), the $\kappa_{\min}\to0$ extrapolated NMEs obtained with the $^{48}$Ca single-reference state and the $^{48}$Ti multi-reference state differ by approximately $\delta_{\rm ref}\simeq0.22$. The particle-hole truncation uncertainty is estimated from the residual convergence pattern in Fig.~\ref{fig:kappa_ref}(b); in particular, the change from the $3p3h$ to the $4p4h$ space in the $^{48}$Ca-reference calculation is about $\delta_{ph}\simeq0.18\text{--}0.20$. Combining these considerations, we adopt a conservative estimate for the lower-end uncertainty of the long-range NME,
$M^{0\nu}_{\rm LR,low}=0.82(20)$. 
We compare this value with the results of other calculations based on the same transition operators, as well as the results by phenomenological nuclear models in Fig.\ref{fig:compare_NME40nu}. It is shown that this value is consistent with the results of the previous ab initio calculations.  The results for the short-range transition operator are shown in Figs.~\ref{fig:CT_kappa_eMax} and \ref{fig:CT_ref}. Based on these calculations, we obtain $M^{0\nu}_{\rm SR}=0.40-0.88$.

For $\Mtwo$, we obtain $M^{2\nu}=0.068~{\rm MeV}^{-1}$ from the calculation using $\eMax=6$  and $\hbar\omega=12$ MeV.  To estimate the uncertainty arising from the truncation in $\eMax$ and importance samplings, we perform calculations using different values of $\eMax$ and different reference states. The results are shown in Fig.~\ref{fig:kappa_2v}. Since the full calculation with larger value of $\eMax$ is challenging, we propagate the uncertainty interval inferred from the lowest-$1^+$ contribution to the full matrix element by shifting the corresponding band to the central value $M^{2\nu}=0.068~{\rm MeV}^{-1}$, as indicated by the black dashed lines in Fig.~\ref{fig:kappa_2v}. This procedure yields 
$M^{2\nu}=0.066\text{--}0.077~{\rm MeV}^{-1}$, which should be interpreted as an approximate truncation uncertainty. This estimate assumes that the dominant lowest-$1^+$ contribution captures the leading threshold dependence of the full matrix element. Contributions from higher-lying intermediate states are not shown explicitly in Fig.~\ref{fig:kappa_2v}; however, they are expected to be less important because they are suppressed by larger energy denominators.

\subsection{The correlation relation analysis with non-implausible chiral Hamiltonians}

\begin{figure}[bt]
   \safeincludegraphics[width=0.8\columnwidth]{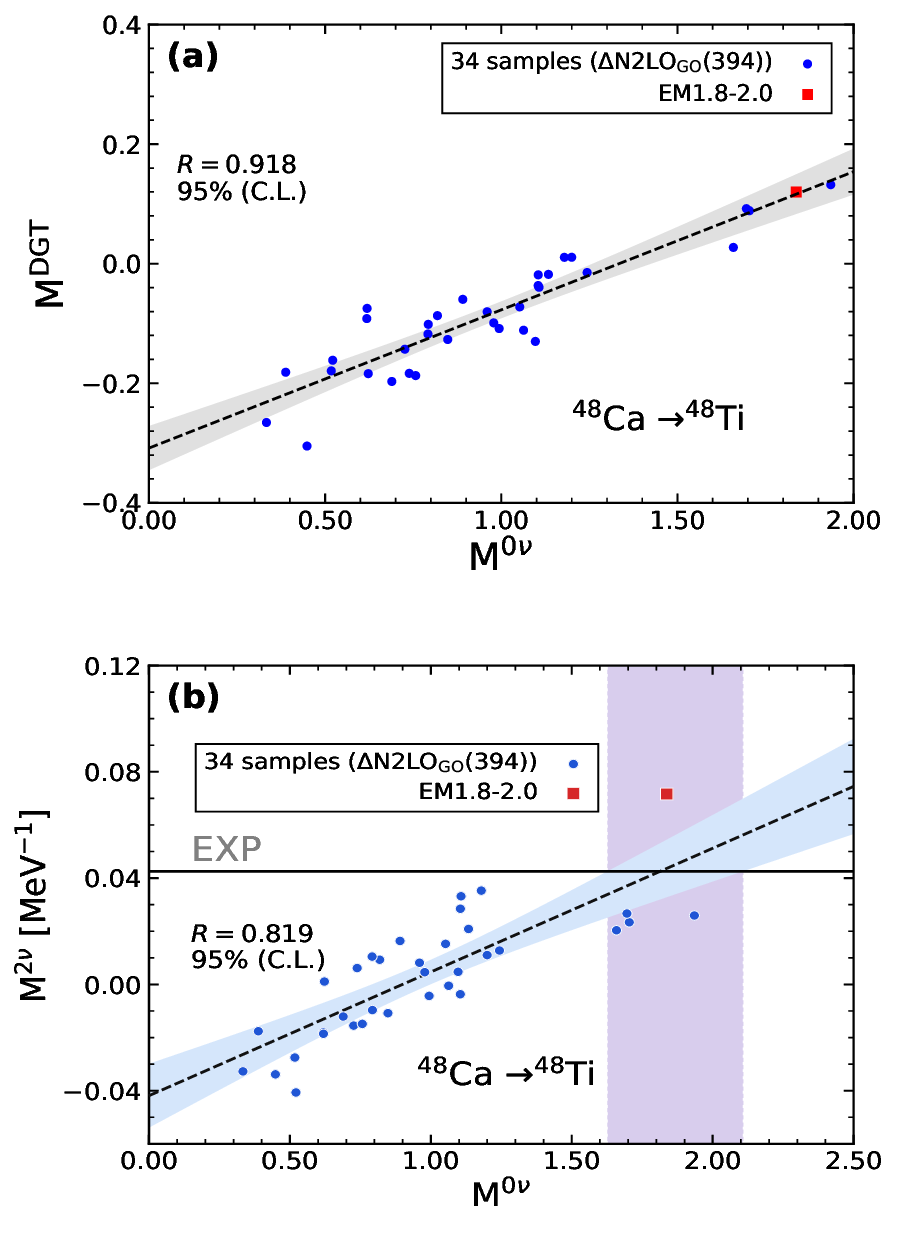}
  \caption{
  Correlations for $^{48}$Ca$\rightarrow{}^{48}$Ti.
  (a) Correlation between the NME of $0\nu\beta\beta$ decay and the DGT matrix element.
  (b) Correlation between the NMEs of $0\nu\beta\beta$ and $2\nu\beta\beta$ decays.
  Blue circles denote the 34 sets of sampled interactions, while the red square marks the EM1.8--2.0 result.
  }
  \label{fig:Correlation_0nu_DGT_2nu_34}
\end{figure}

Figure~\ref{fig:N_extra} shows $M^{2\nu}$ as a function of the excitation energy of the lowest $1^+$ state in $^{48}$Sc, $E_{\rm ex}(1^+_1)$. The two quantities exhibit a weak anti-correlation. Figure~\ref{fig:Correlation_0nu_DGT_2nu_34} displays the correlations between $\Mzero$ and the DGT matrix element, and between $\Mzero$ and $\Mtwo$, for the $^{48}$Ca$\rightarrow{}^{48}$Ti transition, obtained using the 34 non-implausible chiral Hamiltonians from Ref.~\cite{Hu:2022}. In both cases, a strong linear correlation is observed, although the Pearson correlation coefficient for the former is slightly larger than that for the latter. Using the available $2\nu\beta\beta$-decay data together with this correlation relation at the 95\% confidence level, we constrain the total NME to be $M^{0\nu}=1.64$--$2.11$. This value is about 27\% larger than $M^{0\nu}=1.30$--$1.65$, obtained from the correlation relation derived using the 20 non-implausible chiral Hamiltonians. Mathematically, another solution exists, corresponding to a negative value of $M^{2\nu}$. However, the resulting $M^{0\nu}$ is much smaller than the values obtained from ab initio calculations with all the 34 non-implausible chiral Hamiltonians. It is therefore reasonable to discard this solution.

\end{document}